\documentclass[12pt]{article}
\usepackage[english]{babel}
\usepackage[a4paper,top=3 cm,bottom=3 cm,inner=2.5 cm,outer=2.5 cm]{geometry} 
\usepackage[toc,page]{appendix}
\usepackage{graphicx}
\usepackage{titlepic}
\usepackage{bm}
\usepackage{mathtools}
\usepackage{amsmath}
\usepackage{amsfonts}
\usepackage{subcaption}
\usepackage{mathbbol}
\usepackage{enumitem}
\usepackage{cancel}

\setlist[itemize]{label=\textbullet}
\usepackage{float}
\usepackage[table]{xcolor}
\usepackage[utf8]{inputenc}
\usepackage{soul,color}
\usepackage{xcolor}
\usepackage{hyperref}
\usepackage{amsmath, amssymb, amsthm} 
\usepackage{eurosym}
\usepackage{tikz}
\usepackage{csquotes}

\newtheorem{prop}{Proposition} [section]
\newtheorem{defn}{Definition} [section]
\newtheorem{thm}{Theorem}[section] 
\newtheorem{cor}{Corollary}[section]

\theoremstyle{definition}

\usepackage{wasysym}
\hypersetup{pdfborder={0 0 0}} 
\usepackage{url}

\usepackage{titlesec}

\titleformat{\paragraph}
{\normalfont\normalsize\bfseries}{\theparagraph}{1em}{}
\titlespacing*{\paragraph}
{0pt}{3.25ex plus 1ex minus .2ex}{1.5ex plus .2ex}
\usepackage{siunitx}
\usepackage{mathrsfs}

\titleformat{\subparagraph}
{\normalfont\normalsize\bfseries}{\theparagraph}{1em}{}
\titlespacing*{\paragraph}
{0pt}{3.25ex plus 1ex minus .2ex}{1.5ex plus .2ex}
\usepackage{siunitx}
\usepackage{mathrsfs}

\usepackage{leftidx}
\usepackage{braket}
\usepackage{physics}

\usepackage[
backend=biber,
style=numeric-comp,
sorting = none,
isbn = false,
url = false,
maxbibnames = 4,
]{biblatex}
\DeclareFieldFormat[article]{volume}{\mkbibbold{#1}}
\DeclareFieldFormat[article]{number}{\mkbibbold{#1}}
\DeclareFieldFormat[article]{pages}{#1}

\begin{document}
\par 
\bigskip 
\LARGE 
\noindent 
{\bf Relative Entropy of Fermion Excitation States on the CAR Algebra} 
\bigskip \bigskip
\par 
\rm 
\normalsize 
 
\large
\noindent 
{\bf Stefano Galanda$^{1,a}$}, {\bf Albert Much$^{2,b}$}, {\bf Rainer Verch$^{2,c}$} \\
\par
\small

\noindent$^1$ Dipartimento di Matematica, Universit\`a di Genova - Via Dodecaneso, 35, I-16146 Genova, Italy. \smallskip

\noindent$^2$ Institute for Theoretical Physics, University of Leipzig, D-04009 Leipzig, Germany.\smallskip

\smallskip

\noindent E-mail: 
$^a$stefano.galanda@dima.unige.it, 
$^b$much@itp.uni-leipzig.de,
$^c$rainer.verch@uni-leipzig.de\\ 

\normalsize
${}$ \\ \\
 {\bf Abstract} \ \ 
 The relative entropy of certain states on the algebra of canonical anticommutation relations (CAR) is studied in the present work. 
 The CAR algebra is used to describe fermionic degrees of freedom in quantum mechanics and 
 quantum field theory. The states for which the relative entropy is investigated
 are multi-excitation states (similar to multi-particle states) with respect 
 to KMS states defined with respect to a time-evolution induced by a unitary dynamical group on the one-particle Hilbert space of the CAR algebra.
 If the KMS state is quasifree, the relative entropy of multi-excitation states
 can be explicitly calculated in terms of 2-point functions, which are defined 
 entirely by the one-particle Hilbert space defining the CAR algebra and the 
 Hamilton operator of the dynamical group on the one-particle Hilbert space. This applies also in the case that the one-particle Hilbert space Hamilton operator has a continuous spectrum so that the relative entropy of multi-excitation states cannot be defined in terms of von Neumann entropies. The results obtained here
 for the relative entropy of multi-excitation states on the CAR algebra can be viewed as
 counterparts of results for the relative entropy of coherent states on the 
 algebra of canonical commutation relations (CCR) which have appeared recently.
 It turns 
 out to be useful to employ the setting of a self-dual CAR algebra introduced by Araki. 
 \\ \\
${}$

\section{Introduction}

Quantities characterizing the entropy, or change of entropy, have in the recent years gained attention
in quantum field theory, particularly in connection with spacetime geometries of 
black holes and their horizons. We will not review the rich literature on the subject and only 
mention, however selectively, the references \cite{HollandsSanders-book,Mann-book,Solodukhin}; see also further citations therein. The concept
of {\it relative entropy} according to Araki, and Uhlmann \cite{Araki1976,Uhlmann}, between coherent states of a quantized linear Bose field has attracted special interest since it has been found to be a quantity which 
can be calculated explicitly \cite{Hollands-RelEnt,Longo:2019mhx,PhysRevD.99.125020,BosCadVecREccr}. This is especially true in examples related to spacetime horizons
\cite{CiolliEtAl-RelEntCST,HollandsIshibashi-News,PhysRevD.99.125020,KurPinVer,DAngelo21}. Similar considerations for quantized Fermi  fields seem to be missing 
from the literature so far and we set out to fill this gap in the present work. 
\\[6pt]
We briefly describe the result of the explicit expression for the quantized Bose field case that 
has been obtained in \cite{Hollands-RelEnt,Longo:2019mhx,PhysRevD.99.125020,BosCadVecREccr} (using a slightly different notation). It applies, in particular, to coherent states 
of the Weyl-algebra $\mathcal{W}(L,\sigma)$ of a real-linear symplectic space $(L,\sigma)$ ($L$ is 
a real-linear vector space and $\sigma$ a symplectic form on $L$). 
We refer to \cite{Petz:1990,KayWald} for standard properties of the Weyl-algebra, quasifree states etc.\ which we
subsequently summarize.
The Weyl-algebra 
$\mathcal{W}(L,\sigma)$ is the (up to $C^*$-equivalence unique) $C^*$-algebra with
unit element ${\bf 1}$ generated by unitary elements $W(\ell)$, $\ell \in L$, assumed to 
fulfill the relations 
\begin{align}
    W(\ell)W(\ell') = {\rm e}^{-i\sigma(\ell,\ell')/2}W(\ell + \ell') \,,
    \quad W(\ell)^* = W(-\ell) \quad \text{and} \quad W(0) = {\bf 1}\,.
\end{align}
We recall that a state $\omega$ on $\mathcal{W}(L,\sigma)$ is a positive linear functional 
which is positive and unital. 

Starting from a state $\omega$, a {\it coherent state} (relative to $\omega)$ is any state of the form
\begin{align}
    \omega_{\ell}(A) = \omega(W(\ell)AW(\ell)^*)  \quad (A \in \mathcal{W}(L,\sigma))
\end{align}
where $\ell \in L$ is arbitrary. 
\par
Now also assume that there is a (smooth) one-parametric group of symplectomorphisms $\{\Lambda_t\}_{t \in \mathbb{R}}$ on $L$ inducing a 1-parametric group of automorphisms $\{\lambda_t\}_{t \in \mathbb{R}}$
on $\mathcal{W}(L,\sigma)$ by $\lambda_t(W(\ell)) = W(\Lambda_t \ell)$  and that the quasifree state $\omega$ is a KMS-state at inverse temperature $\beta = 1$
for the automorphism group $\{\lambda_t\}_{t \in \mathbb{R}}$. In this case, the result found 
in the above cited references states that the Araki-Uhlmann relative entropy $S(\omega_\ell \| \omega)$
of the coherent state $\omega_\ell$ with respect to the state $\omega$ is given by 
\begin{equation} \label{eq:coherentent}
   S(\omega_\ell \| \omega) = \left. \frac{1}{2} \frac{d}{dt} \right|_{t = 0} \sigma(\Lambda_t \ell, \ell) \,, \quad \ell \in L\,.
\end{equation} 
As mentioned, the right hand side is amenable to explicit calculation in many scenarios in quantum
field theory, especially in the presence of spacetime horizons, and has shown its utility in that
context \cite{CiolliEtAl-RelEntCST,HollandsIshibashi-News,PhysRevD.99.125020,KurPinVer,DAngelo21}. This possiblity of explicit calculation makes the relative entropy an attractive
quantity in quantum field theory. At first sight that may come unexpected as the general definition of relative entropy 
between states of operator algebras is abstract, relying  on Tomita-Takesaki modular objects, as 
we will explain in more detail in the following section. 
\par
That explained, a similar formula for the relative entropy in the Fermi case, i.e.\ for 
states on the CAR-algebra of the {\it canonical anti-commutation relations}, appears to be
absent from the literature. The reason may be, among others, that the concept of a coherent
state doesn't lend itself as naturally for the CAR-algebra as it does for the Weyl-algebra. 
However, as we   show, using the {\it self-dual} approach to the CAR-algebra of Araki 
\cite{ArakiQuasifreeCAR} a very natural notion of {\it (multi-)excited} states with respect to a KMS state can be given,
and if the KMS flow is induced by a unitary 1-parametric group on the one-particle Hilbert space from
which the CAR-algebra is generated, then again a very simple formula for the relative entropy 
of an excited state with respect to that KMS state can be obtained. Moreover, the 
relative entropy of (multi-)excited states on the CAR-algebra can be calculated 
very similarly as in the case of the Weyl-algebra. To be a bit more explicit (full details will be given in the main text), Araki's self-dual CAR algebra, denoted by {\tt CAR}$(\mathcal{K},\Gamma)$ is defined with respect to a complex Hilbert space 
$\mathcal{K}$, called the one-particle Hilbert space, equipped with a complex
conjugation $\Gamma$, as the (unique) $C^*$-algebra generated by a unit element
${\bf 1}$ and elements $B({\sf f})$, ${\sf f} \in \mathcal{K}$. The generating elements are assumed to be linear in ${\sf f}$ and   fulfill the relations
\begin{align}
 B({\sf f})^* = B(\Gamma {\sf f}) \quad \text{and} \quad B({\sf f})^*B({\sf g}) + B({\sf g})B({\sf f})^* = \braket{{\sf f}}{{\sf g}}_{\mathcal{K}} {\bf 1},
\end{align}
where the bracket denotes the scalar product of the complex Hilbert space $\mathcal{K}$. The interesting
point is that therefore, if $\Gamma {\sf f} = {\sf f}$ and $\braket{{\sf f}}{{\sf f}}_{\mathcal{K}} = 2$, $B({\sf f})$ is unitary (and, at the same time, hermitean). This means, if $\omega$ is a standard state on 
{\tt CAR}$(\mathcal{K},\Gamma)$, i.e.\ it extends to a faithful state 
on the induced von Neumann algebra in its GNS representation, then
for the state $\omega_{\sf f}(\,.\,) = \omega(B({\sf f})^*\,.\,B({\sf f}))$,
the relative entropy $S(\omega_{\sf f}\|\omega)$ can be defined following the 
definition given by Araki \cite{Araki1976}. 
As we   show in Theorem \ref{thm:AandB},
if there is a continuous unitary group $\{u_t\}_{t \in \mathbb{R}}$ on 
$\mathcal{K}$ which induces the modular group of $\omega$ for its induced 
von Neumann algebra in the GNS representation, then the relative entropy 
is given by 
\begin{align}
 S(\omega_{\sf f}\|\omega) =  \braket{{\sf f}}{Q_{(1)}\boldsymbol{h}{\sf f}}_{\mathcal{K}} 
\end{align}
where $\boldsymbol{h}$ is the selfadjoint generator of 
$\{u_t\}_{t \in \mathbb{R}}$ and $Q_{(1)} = ({\bf 1} + {\rm e}^{-\boldsymbol{h}})^{-1}$ is defined by 
$\boldsymbol{h}$ through the spectral calculus. These results rely to large 
part on results obtained by Araki on states of the self-dual CAR algebra 
\cite{ArakiQuasifreeCAR}. 
\par
Moreover, multi-excitation states can also be defined with respect to a state $\omega$ with properties as before in a similar way,
replacing $B({\sf f})$ by a product $B({\sf f}_1) \cdots B({\sf f}_n)$. If $\omega$ is quasifree, i.e.\ determined by its 2-point function, then a compact formula for the relative entropy of the
multi-excitation state with respect to $\omega$ can be established. Thus, the (multi-)excitation states of the CAR algebra play an analogous role to the coherent states of the CCR algebra when studying relative entropy. The utility of Araki's self-dual approach to the CAR algebra is also 
demonstrated in that context. 
\par
This article is organized as follows. In Section 2, we summarize the definition of 
the self-dual CAR algebra according to Araki \cite{ArakiQuasifreeCAR}, as well as the definition
of relative entropy in a general von Neumann algebra setting. In the same section  we introduce 
single- and multi-excitation states relative to some state $\omega$ which is taken to be 
a KMS state with respect to a dynamics induced by a unitary group on the one-particle space,
and present some results on their relative entropies. In Section 3, we focus on the case that
$\omega$ is a quasifree state, and we obtain further results on the relative entropy for multi-excitation states. Section 4 demonstrates that the relative entropy of multi-excitation states
studied so far coincides with the relative entropy given in terms of von Neumann entropies whenever
the relevant density matrices are available. 
In Section 5, we give a brief summary and conclusion. 
In the Appendix, we briefly recapitulate the definition of ground- and KMS states on the CAR algebra.
\par
The material presented in this article has partially been taken (and generalized) from the first named author's MSc thesis \cite{Galanda} where also some applications can be found.

\section{Self-dual CAR-algebra and relative entropy}

\subsection{Self-dual CAR algebra: generalities} 
Let $\mathcal{K}$ be a complex Hilbert space with scalar product 
$\braket{\cdot}{\cdot}_{\mathcal{K}}$, together with an antilinear operator $\Gamma : \mathcal{K} \to
\mathcal{K}$ which is antiunitary, so that 
\begin{align}
     \Gamma^2 = {\bf 1} \quad \text{and} \quad 
    \braket{\Gamma {\sf f}}{\Gamma {\sf g}}_{\mathcal{K}} = \braket{{\sf g}}{{\sf f}}_{\mathcal{K}} \quad \ \
    ({\sf f},{\sf g} \in \mathcal{K})
\end{align}

\begin{defn}
{\rm  \cite{ArakiQuasifreeCAR} } \ \ {\tt CAR}$(\mathcal{K},\Gamma)$, the $C^*$-algebra of the {\rm self-dual canonical anti-commu\-tation 
relations} with respect to $(\mathcal{K},\Gamma)$, is defined as the (up to $C^*$-equivalence unique)
unital $C^*$-algebra generated by the unit element ${\bf 1}$ and elements $B({\sf f})$, ${\sf f} \in 
\mathcal{K}$, fulfilling the conditions
\begin{align}
   & {\sf f} \mapsto B({\sf f})  \ \ \text{is complex linear}\,, \quad B({\sf f})^* = B(\Gamma {\sf f})\,, \quad \text{and}\\
   &  [B({\sf f})^*, B({\sf g})]_+  = \braket{{\sf f}}{{\sf g}}_{\mathcal{K}} {\bf 1}
\end{align}
for all ${\sf f},{\sf g} \in \mathcal{K}$, where $[X,Y]_+ = XY + YX$ denotes the algebraic
{\it anti-commutator bracket}.
\end{defn}

For the implicit statement that {\tt CAR}$(\mathcal{K},\Gamma)$ indeed forms a $C^*$-algebra and 
its uniqueness (up to $C^*$-equivalence), see \cite{ArakiQuasifreeCAR}. 
We often call $\mathcal{K}$ the {\it one-particle space} of {\tt CAR}$(\mathcal{K},\Gamma)$.
\\[6pt]
Of interest is the case that unitaries on the one-particle space induce $C^*$-automorphisms on
{\tt CAR}$(\mathcal{K},\Gamma)$, and we formulate it a follows:

\begin{prop}\label{prop: dynamics} {\rm  \cite{ArakiQuasifreeCAR}} Let $\{u_t\}_{t \in \mathbb{R}}$ be 
a 1-parametric continuous unitary group on $\mathcal{K}$ with the property that 
\begin{align}\label{eq:1Pflow} \Gamma u_t = u_t \Gamma \quad \ \ (t \in \mathbb{R})\,.
    \end{align}
Then there is a 1-parametric continuous group $\{{\tt T}_t\}_{t \in \mathbb{R}}$ of unital automorphisms
of {\tt CAR}$(\mathcal{K},\Gamma)$ defined by 
\begin{align} \label{eq:induced-flow}
    {\tt T}_t(B({\sf f})) = B(u_t{\sf f}) \quad \ \  ({\sf f} \in \mathcal{K}\,,\ t \in \mathbb{R})\,.
\end{align}
\end{prop}

In this case we say that the automorphism group $\{{\tt T}_t\}_{t \in \mathbb{R}}$ is 
induced by the {\it 1-particle space flow} $\{u_t\}_{t \in \mathbb{R}}$ or simply by a {\it 1-particle flow}, for short.
\\[6pt]
We need to recall some standard terminology in the context of $C^*$-algebras, see, e.g., \cite{BraRob1} for further details and proofs of statements that will be reproduced below. Let $\mathcal{C}$ be 
a $C^*$-algebra with unit element ${\bf 1}$  (we may think, in particular, of $\mathcal{C} =$ {\tt CAR}$(\mathcal{K},\Gamma)$).
A {\it state} $\omega$ on $\mathcal{C}$ is a linear functional $\omega: \mathcal{C} \to \mathbb{C}$
which is positive, i.e.\ $\omega(C^*C) \ge 0$ for all $C \in \mathcal{C}$, and normalized, i.e.\
$\omega({\bf 1}) = 1$. For any given state $\omega$ on $\mathcal{C}$, there is an associated 
Hilbert space representation together with a distinguished (up to phase, unique) unit vector, denoted by
$(\mathcal{H}_\omega,\pi_\omega,\Omega_\omega)$, where $\mathcal{H}_\omega$ is the representation
Hilbert space, $\pi_\omega$ is the unital $*$-representation of $\mathcal{C}$ by bounded linear 
operators on $\mathcal{H}_\omega$, and $\Omega_\omega$ is the said distinguished unit vector. It
is distinguished by the following properties: \ \  (I) $\omega(C) = (\Omega_\omega,\pi_\omega(C)\Omega_\omega)$
for all $C \in \mathbb{C}$, where the round brackets denote the scalar product of $\mathcal{H}_\omega$.
 \ \ (II) $\Omega_\omega$ is cyclic for the represented $C^*$-algebra $\pi_\omega(C)$, that is, the 
vector space formed by all $\pi_\omega(C)\Omega_\omega$, with $C$ ranging over all elements in $\mathcal{C}$, is dense in $\mathcal{H}_\omega$. The triplet 
$(\mathcal{H}_\omega,\pi_\omega,\Omega_\omega)$ is 
called the {\it GNS-representation} of $\omega$. 

In the GNS representation of $\omega$, one can form the weak closure of 
the represented $C^*$-algebra, thereby obtaining the {\it induced von Neumann algebra} $\mathcal{N}_\omega = \overline{\pi_\omega(\mathcal{C})}^{\rm weak}$ on the Hilbert space 
$\mathcal{H}_\omega$. The general concept of relative entropy refers to such a von Neumann algebra setting. Moreover, it is worth noting that the state
$\omega$ on $\mathcal{C}$ extends naturally to a state on the induced von Neumann 
algebra $\mathcal{N}_\omega$, as the state induced by the state vector $\Omega_\omega$;
therefore that extension of $\omega$ (denoted by the same symbol) is given by 
\begin{align}
    \omega(A) = (\Omega_\omega,A\Omega_\omega) \ \ \quad (A \in \mathcal{N}_\omega)\,.
\end{align}
Clearly, $\Omega_\omega$ is cyclic for $\mathcal{N}_\omega$, so that $\mathcal{N}_\omega\Omega_\omega$
is a dense subspace of $\mathcal{H}_\omega$. Then $\Omega_\omega$ is, equivalently, separating for 
the commutant von Neumann algebra $\mathcal{N}'_\omega = \{A \in \mathcal{B}(\mathcal{H}_\omega): AN = NA \quad \text{for all} \ N \in \mathcal{N}_\omega\}$ and this means that $A\Omega_\omega = 0$ is only 
possible if $A = 0$ for any $A \in \mathcal{N}'_\omega$.

\subsection{Relative entropy: general definition}

Let $\mathcal{H}$ be a Hilbert space and let $\mathcal{N} \subset \mathcal{B}(\mathcal{H})$ be a von Neumann
algebra. We assume that there are two unit vectors $\Omega$ and $\Psi$ in $\mathcal{H}$ which are 
both cyclic and separating for $\mathcal{N}$. Then, there is the {\it relative Tomita operator}
$S_{\Psi,\Omega}: A\Omega \mapsto A^*\Psi$, $A \in \mathcal{N}$. The operator is defined on the 
dense domain $\mathcal{N}\Omega$; it is conjugate-linear and closable and its closure
(again denoted by $S_{\Psi,\Omega}$) admits a (unique)
polar decomposition $S_{\Psi,\Omega} = J_{\Psi,\Omega}\Delta^{1/2}_{\Psi,\Omega}$ where $J_{\Psi,\Omega}$ is an 
anti-linear conjugation on $\mathcal{H}$ (called {\it relative modular conjugation}) and $\Delta^{1/2}_{\Psi,\Omega}$ is a closed, positive operator (its square $\Delta_{\Psi,\Omega}$ is 
called {\it relative modular operator}). The two unit vectors $\Omega$ and $\Psi$ induce 
faithful, normal states on $\mathcal{N}$:
\begin{align}
    \omega_\Omega(A) = (\Omega,A\Omega)\,, \quad \omega_\Psi(A) = (\Psi,A\Psi)\ \ \quad (A \in \mathcal{N})\,,
\end{align}
where the paired round brackets denote the scalar product of $\mathcal{H}$.
Following Araki (and independently, Uhlmann) \cite{Araki1976,Uhlmann}, the {\it relative entropy of the state $\omega_\Psi$ with respect to the state $\omega_\Omega$} is defined as 
\begin{align} \label{eq:defrelent}
    S(\omega_\Psi \| \omega_\Omega) = - (\Omega,\log(\Delta_{\Psi,\Omega}) \Omega) = i
    \left. \frac{d}{dt} \right|_{t = 0} (\Omega,\Delta^{it}_{\Psi,\Omega} \Omega)\,,
\end{align}
provided that the derivative on the right hand side exists, in which case it is 
a finite, non-negative number. It is customary to formally define $S(\omega_\Psi \| \omega_\Omega) = +\infty$ in the case that the derivative on the right hand side 
does not exist. 
The definition applies more generally for any pair of faithful normal states $\omega = \omega_\Omega$
and $\tilde{\omega} = \omega_\Psi$ on $\mathcal{N}$ independent of the choice of their implementing unit vectors (here: $\Omega$ and $\Psi$). For further discussion of the properties of the thus defined 
relative entropy, its interpretation, and generalizations, we refer to the references 
\cite{Araki1976,Uhlmann,OhyaPetz-book}. 

We put on record some known results about the properties of relative entropy, for later use;
recall that for any unit vector $\Omega$ in $\mathcal{H}$ which is cyclic and separating for 
$\mathcal{N}$, and any unitary operator $U \in \mathcal{N}$, the vector $U\Omega$ is again a cyclic and separating unit vector for $\mathcal{N}$. 

\begin{prop}
 {\rm (1)} Let $\alpha : \mathcal{N} \to \mathcal{N}$ be a von Neumann algebra automorphism. Then,
 \begin{align} \label{eq:reentrafo}
  S(\omega_\Psi \circ \alpha \| \omega_\Omega \circ \alpha) = S(\omega_\Psi \| \omega_\Omega) \,.
 \end{align}
Again, this generalizes to faithful, normal states $\omega$ and $\tilde{\omega}$ in place of $\omega_\Omega$ and 
$\omega_\Psi$, respectively. 
\\ \\
{\rm (2)}  Let $U\in \mathcal{N}$ be unitary and let $\Psi = U\Omega$. Then
\begin{align} \label{eq:relent}
 S(\omega_\Psi \| \omega_\Omega) =    S(\omega_{U\Omega} \| \omega_\Omega) = i
    \left. \frac{d}{dt} \right|_{t = 0} (U^*\Omega,\Delta^{it}_{\Omega,\Omega} U^*\Omega)\,,
\end{align}
hence, $S(\omega_\Psi \| \omega_\Omega)$ is finite if $U^*\Omega$ lies in the 
domain of $|\log(\Delta_{\Omega,\Omega})|^{1/2}$.
\end{prop}
\begin{proof}
For proof   of   statement $(1)$ see \cite[Equation (6.6)]{Araki1976}  and for a proof of statement $(2)$ see \cite[Lemma 5.7]{PhysRevD.99.125020}.
\end{proof}

\subsection{Relative entropy for excitations with respect to a standard state on the CAR algebra}

Let $\omega$ be a state on {\tt CAR}$(\mathcal{K},\Gamma)$. We call such a state {\it standard} if the GNS-vector $\Omega_\omega$ of the GNS representation $(\mathcal{H}_\omega,\pi_\omega,\Omega_\omega)$ is cyclic and separating for the induced von Neumann algebra
$\mathcal{N}_\omega = \overline{\pi_\omega({\tt CAR}(\mathcal{K},\Gamma))}^{\rm weak}$. 
Consequently, one can associate Tomita-Takesaki modular objects with the pair 
$(\mathcal{N}_\omega,\Omega_\omega)$, in particular the {\it modular operator} $\Delta$ and 
the {\it modular group} $\{ \Delta^{it} \}_{t \in \mathbb{R}}$. Note that $\Delta =
\Delta_{\Omega_\omega,\Omega_\omega}$ in our previously used notation, but the notation
with the subscripts is cumbersome and we drop it as no confusion is likely to arise. 
\\[6pt] 
We observe that for ${\sf f} \in \mathcal{K}$ fulfilling $\Gamma{\sf f} = {\sf f}$ and 
$\braket{{\sf f}}{{\sf f}}_{\mathcal{K}} = 2$ the CAR-generator $B({\sf f})$ is unitary
since, as a consequence of the defining relations of the CAR-generators, it holds for 
the ${\sf f}$ with the said properties that 
\begin{align}
    B({\sf f})^* = B(\Gamma{\sf f}) = B({\sf f})\,, \quad \text{implying} \quad 
    2B({\sf f})B({\sf f}) = \braket{{\sf f}}{{\sf f}}_{\mathcal{K}}{\bf 1} =  2 \cdot {\bf 1}
\end{align}
from which one concludes $B({\sf f})^*B({\sf f}) = B({\sf f})B({\sf f})^* = {\bf 1}$. 
In consequence, for any such ${\sf f}$, its GNS-represented counterpart 
$F = \pi_\omega(B({\sf f}))$ is a unitary operator contained in $\mathcal{N}_\omega$. 

\begin{defn} \label{def:stand}
    Let $\omega$ be a standard state on {\tt CAR}$(\mathcal{K},\Gamma)$.
\\[6pt]
(a) \ \ For any ${\sf f}$ with $\Gamma{\sf f} = {\sf f}$ and 
$\braket{{\sf f}}{{\sf f}}_{\mathcal{K}} = 2$, we call the state on $\mathcal{N}_\omega$
induced by the state vector $F\Omega_\omega$, where $F = \pi_\omega(B({\sf f}))$, 
a  \emph{single-excitation state} relative to $\omega$, and we denote it by $\omega_{\sf f}$.
\\[6pt]
(b) \ \
Let ${\sf f}_1,\ldots,{\sf f}_n$
 be finitely many elements in $\mathcal{K}$ with $\Gamma{\sf f}_j = {\sf f}_j$
 and $\braket{{\sf f}_j}{{\sf f}_j}_{\mathcal{K}} = 2$ $(j = 1,\ldots,n)$, so that in the GNS-representation
 of $\omega$, the operators $F_j = \pi_\omega(B({\sf f}_j))$ are unitary. 
 Then, we call the standard state $\omega_{{\sf f}_1\cdots{\sf f}_n}$ on 
 $\mathcal{N}_\omega$
 given by 
 \begin{align} 
  \omega_{{\sf f}_1\cdots{\sf f}_n}(A) = (F_1 \cdots F_n\Omega_\omega,A F_1 \cdots F_n\Omega_\omega)
  \quad \ \ (A \in \mathcal{N}_\omega)
 \end{align}
a \emph{multi-excitation state} relative to $\omega$.
\\[6pt]
(c) \ \ We call a standard state $\omega$ a \emph{standard 1-particle flow state (S1PFS)} if there 
is a 1-particle flow $\{u_t\}_{t \in \mathbb{R}}$ on $(\mathcal{K},\Gamma)$ (i.e.\ a 1-parametric
unitary group with the property \eqref{eq:1Pflow}) such that 
\begin{align} \label{eq:flowinduced}
 \Delta^{it} \pi_\omega(B({\sf f})) \Delta^{-it} = \pi_\omega(B(u_t{\sf f}))
\end{align}
holds for all ${\sf f} \in \mathcal{K}$ and all $t \in \mathbb{R}$. 
    
\end{defn}
We observe that a standard 1-particle flow state is, equivalently, a KMS-state at inverse 
temperature $\beta = 1$ on {\tt CAR}$(\mathcal{K},\Gamma)$ with respect to the 
automorphism group $\{{\tt T}_t\}_{t \in \mathbb{R}}$ induced by the 1-particle space flow
$\{u_t\}_{t\in\mathbb{R}}$ as in Proposition  \ref{prop: dynamics}. (The definition of KMS state is summarized in the Appendix.) This is the statement
of Takesaki's theorem, cf.\
Theorem  5.3.10 in \cite{BraRob2}.  Furthermore, if $\omega$ is a standard state, then so is any
of its single/multi-excitation states
$\omega_{\sf f}$. 
\\ \\
Before stating the main result of this section, it is useful to introduce some further 
terminology, following \cite{ArakiQuasifreeCAR} to large extent. 

For any state $\omega$ on {\tt CAR}$(\mathcal{K},\Gamma)$, we define its {\it 2-point function}
\begin{align}
 W^{(2)}_\omega({\sf f},{\sf g}) = \omega(B({\sf f})B({\sf g})) \quad \ \ ({\sf f},{\sf g} \in \mathcal{K})\,.
\end{align}
By the properties of a state, it follows that ${\sf f},{\sf g} \mapsto W^{(2)}_\omega(\Gamma{\sf f},{\sf g})$ is a sesquilinear form on $\mathcal{K}$ dominated by the scalar product $\langle \,.\,|\,.\,\rangle_{\mathcal{K}}$, from which it can be concluded (cf.\ \cite{ArakiQuasifreeCAR})
that there is for any state $\omega$ on {\tt CAR}$(\mathcal{K},\Gamma)$ a linear operator 
$Q_\omega : \mathcal{K} \to \mathcal{K}$, called the {\it base polarization} of $\omega$, 
characterized by the properties:
\begin{align}
 0 \le Q_\omega = &\ Q^*_\omega \le {\bf 1}\,,  \quad Q_\omega + \Gamma Q_\omega \Gamma = {\bf 1}\,, \quad \text{and} \\
 & W^{(2)}_\omega(\Gamma{\sf f},{\sf g}) = \braket{{\sf f}}{Q_\omega{\sf g}}_{\mathcal{K}}.
\end{align}
If $\omega$ is a standard state, then $W^{(2)}_\omega$ is non-degenerate and the sesquilinear form that it
induces is a scalar product on $\mathcal{K}$. For a S1PFS $\omega$, one can specifically state the form
of $Q_\omega$. 
To this end, we assume a 1-particle space flow 
 $\{u_t\}_{t \in \mathbb{R}}$ on
$\mathcal{K}$ is given.
Then $u_t = {\rm e}^{-it\boldsymbol{h}}$ with a selfadjoint operator $\boldsymbol{h}$
defined on a dense domain ${\rm dom}(\boldsymbol{h})$ in $\mathcal{K}$. 
Since $\Gamma$ is anti-unitary, the commutativity of $\Gamma$ and $u_t$ implies
that $\Gamma \boldsymbol{h} = - \boldsymbol{h} \Gamma$. In turn, this means that 
${\rm spec}(\boldsymbol{h})$, the spectrum of $\boldsymbol{h}$, is 
invariant under the reflection $\lambda \mapsto - \lambda$ on the real axis. Now for 
$\beta > 0$, we define
\begin{align} \label{eq:Qbeta}
 Q_{(\beta)} = ({\bf 1} + {\rm e}^{-\beta \boldsymbol{h}})^{-1}
\end{align}
by the spectral calculus. On citing the following result, see {\rm \cite{ArakiQuasifreeCAR}}, note that the definition of 
{\it quasifree state} on {\tt CAR}$(\mathcal{K},\Gamma)$ appearing in the statement will
be given in the next section. 
\begin{prop}
 Suppose that the automorphism group $\{{\tt T}_t\}_{t \in \mathbb{R}}$ on
 {\tt CAR}$(\mathcal{K},\Gamma)$ is induced by a 1-particle flow $\{u_t\}_{t \in \mathbb{R}}$ as in
 \eqref{eq:induced-flow}.
 \\[4pt]
{\rm (A)} If $\omega$ is a KMS state for $\{{\tt T}_t\}_{t \in \mathbb{R}}$ at inverse temperature 
$\beta > 0$, the 2-point function of $\omega$ is given by
\begin{align} \label{eq:KMS}
  W_\omega^{(2)}({\sf f},{\sf g}) = 
  \braket{\Gamma{\sf f}}{Q_{(\beta)}{\sf g}}_{\mathcal{K}}
  \quad ({\sf f},{\sf g} \in \mathcal{K})
 \end{align}
{\rm (B)} Conversely, for any inverse temperature $\beta > 0$ there is a  \emph{quasifree} KMS state $\omega$ for 
$\{{\tt T}_t\}_{t \in \mathbb{R}}$
which is defined by the 2-point function given 
\eqref{eq:KMS}, and by defining the $n$-point functions of the 
respective states as in \eqref{eq:np-odd} and \eqref{eq:np-even} in the definition of 
quasifree states appearing below.
\\[4pt]
{\rm (C)} If $\omega$ is a KMS state for $\{{\tt T}_t\}_{t \in \mathbb{R}}$ at inverse temperature $\beta > 0$ such 
that the induced von Neumann algebra $\mathcal{N}_\omega$ in the GNS representation
of $\omega$ is a factor, i.e.\ $\mathcal{N}_\omega \cap \mathcal{N}_\omega' = \mathbb{C}{\bf 1}$, then $\omega$ coincides with the quasifree state determined
by the 2-point function \eqref{eq:KMS}. 
\\[4pt]
The following conditions imply that $\mathcal{N}_\omega$ is a factor:
(1) $1/2$ is not an eigenvalue of $Q_{(\beta)}$, or (2) $1/2$ is an eigenvalue of 
$Q_{(\beta)}$ with finite and even multiplicity, or (3) $1/2$ is an eigenvalue 
of $Q_{(\beta)}$ with infinite multiplicity. 
\end{prop}
\begin{cor}
 {\rm (i)} \ \ If $\omega$ is a {\rm S1PFS}, then its 2-point function has the form 
\begin{align}
 W_\omega^{(2)}({\sf f},{\sf g}) & = \braket{\Gamma{\sf f}}{Q_{(1)}{\sf g}}_{\mathcal{K}} \quad \ \ 
 ({\sf f},{\sf g} \in \mathcal{K}) \ \ \text{where} \label{eq:unique}\\
 Q_{(1)}  & = ({\bf 1} + {\rm e}^{-\boldsymbol{h}})^{-1} \label{eq:theQ1}
\end{align}
{\rm (ii)} \ \ The condition that $1/2$ is not an eigenvalue of $Q_{(1)}$ is equivalent to the 
statement that $0$ is not an eigenvalue of $\boldsymbol{h}$. The latter is often referred to as 
the condition of ``absence of zero modes for $\boldsymbol{h}$''. Therefore, if the condition of 
``absence of zero modes for $\boldsymbol{h}$'' holds, then the {\rm S1PFS} is unique: It is the 
quasifree state defined by the 2-point function \eqref{eq:unique}, or equivalently, by the base
polarizator $Q_{(1)}$. 
\end{cor}
After these preparations, we can state one of our central results.
\begin{thm} \label{thm:AandB} ${}$
\\[6pt]
{\rm (A)} \ \
 Let $\omega$ be a {\rm S1PFS} on {\tt CAR}$(\mathcal{K},\Gamma)$, where 
 $\{u_t\}_{t \in \mathbb{R}}$ with $u_t = {\rm e}^{-it\boldsymbol{h}}$ denotes the associated 1-particle flow. Then, for any single-excitation state $\omega_{\sf f}$ relative to $\omega$, the relative entropy is 
 \begin{align} \label{eq:1exent}
  S(\omega_{\sf f} \| \omega) & = i
    \left. \frac{d}{dt} \right|_{t = 0} W^{(2)}_\omega({\sf f},u_t{\sf f}) \\
    &=   \braket{{\sf f}}{Q_{(1)}\boldsymbol{h}{\sf f}}_{\mathcal{K}} 
\end{align}
where $Q_{(1)}$ is given by \eqref{eq:theQ1}.
Thus, the relative entropy $S(\omega_{\sf f}\|\omega)$ is finite if 
${\sf f}$ is contained in the 
domain of $|\boldsymbol{h}|^{1/2}$.
\\[6pt]
{\rm (B)} \ \
Let $\omega_{{\sf f}_1\cdots {\sf f}_n}$ and $\omega_{{\sf g}_1\cdots{\sf g}_m}$ be 
 multi-excitation states relative to a standard state $\omega$ on {\tt CAR}$(\mathcal{K},\Gamma)$. 
 Then, 
 \begin{align}
  S(\omega_{{\sf f}_1 \cdots {\sf f}_n} \| \omega_{{\sf g}_1 \cdots {\sf g}_m}) = 
  S(\omega_{{\sf g}_m \cdots {\sf g}_1 {\sf f}_1 \cdots {\sf f}_n} \| \omega).
 \end{align}
\end{thm}
{\it Proof}. (A) \ \ With the notation as in Definition \ref{def:stand}, we obtain by Equation \eqref{eq:relent} for the relative entropy
\begin{align} \label{eq:rent}
S(\omega_{\sf f} \| \omega) & = i
    \left. \frac{d}{dt} \right|_{t = 0} (\Omega_\omega,F \Delta^{it} F\Omega_\omega)
\end{align}
and the expressions for the right hand side of the equations in the statement of the Theorem are just re-writings 
of the right hand side of Equation \eqref{eq:rent}. 
\\[6pt]
(B) \ \ 
We set $F_j = \pi_\omega(B({\sf f}_j))$ and $G_k = \pi_\omega(B({\sf g}_k))$, where 
$j = 1,\ldots,n$ and $k = 1,\ldots,m$.
Defining for any unitary $W \in \mathcal{N}_\omega$ the automorphism $\alpha_W(A) = W^*AW$ $(A \in \mathcal{N}_\omega)$ of $\mathcal{N}_\omega$, we obtain, with the unitaries $U = F_1 \cdots F_n$ and
$V = G_1 \cdots G_m$, 
\begin{align}
\omega_{{\sf f}_1 \cdots {\sf f}_n} = \omega \circ \alpha_U \quad \text{and} \quad 
\omega_{{\sf g}_1 \cdots {\sf g}_m} = \omega \circ \alpha_V\,.
\end{align}
On using Relation \eqref{eq:reentrafo}, it holds that $S(\omega \circ \alpha_U \| \omega \circ \alpha_V) = S(\omega \circ \alpha_U \circ \alpha_V^{-1} \| \omega)$. We observe that 
$\alpha_U \circ \alpha_V^{-1} = \alpha_{V^*U}$, whence we arrive at 
\begin{align}
 S(\omega_{{\sf f}_1 \cdots {\sf f}_n} \| \omega_{{\sf g}_1 \cdots {\sf g}_m}) & = 
 S(\omega \circ \alpha_U \| \omega \circ \alpha_V) \\
 & = S(\omega \circ \alpha_{V^*U}\| \omega) =
  S(\omega_{{\sf g}_m \cdots {\sf g}_1 {\sf f}_1 \cdots {\sf f}_n} \| \omega)\,.
\end{align}
${}$ \hfill $\Box$
\\ \\
We remark that Formula \eqref{eq:rent} gives the relative entropy between a single-excitation state $\omega_{\sf f}$ and $\omega$ in general, without assuming that the modular group $\{\Delta^{it}\}_{t \in \mathbb{R}}$ of 
$(\mathcal{N}_\omega,\Omega_\omega)$ is induced by a 1-particle flow as in Equation \eqref{eq:flowinduced}.

\subsection{Relative entropy for unitary exponential excitations}

For the single particle vectors ${\sf f} \in \mathcal{K}$ fulfilling $\Gamma {\sf f} = {\sf f}$ and $\braket{{\sf f}}{{\sf f}}_{\mathcal{K}} = 2$, it holds that $B({\sf f}) = B({\sf f})^*$, implying that the exponential series ${\rm e}^{iB({\sf f})} = \sum_{n = 0}^\infty 
i^n B({\sf f})^n/n!$ is a unitary element in {\tt CAR}$(\mathcal{K},\Gamma)$. 
Making use of the fact that $B({\sf f})^2 = {\bf 1}$, one finds
\begin{align}
    {\rm e}^{i B({\sf f})} &= {\bf 1} + i B({\sf f}) +\frac{(i)^2}{2!}B({\sf f})^2 + \frac{(i)^3}{3!}B({\sf f})^3 + \dots\\
    &= {\bf 1} + i B({\sf f}) +\frac{(i)^2}{2!} {\bf 1} + \frac{(i)^3}{3!}B({\sf f}) + \dots\\
    &= \bigg( 1 - \frac{1}{2!} +  \frac{1}{4!} - \dots\bigg) {\bf 1} + i \bigg(1 - \frac{1}{3!} + \frac{1}{5!} - \dots\bigg) B({\sf f})\\
    &= \cos(1) {\bf 1} + i \sin(1) B({\sf f})\,.
\end{align}
Hence, if $\omega$ is a standard state on {\tt CAR}$(\mathcal{K},\Gamma)$, with GNS
representation $(\mathcal{H}_\omega,\pi_\omega,\Omega_\omega)$ as before, the exponentiated
unitary $E_{\sf f} = \pi_\omega({\rm e}^{iB({\sf f})})$ in the GNS representation takes the form
\begin{align}
 E_{\sf f} = \cos(1){\bf 1} + i \sin(1) F\,, \quad F = \pi_\omega(B({\sf f}))\,.
\end{align}
On the von Neumann algebra $\mathcal{N}_\omega$, defined as before, we can define 
the state 
\begin{align}
 \omega_{E_{\sf f}\Omega_\omega}(A) = (E_{\sf f}\Omega_\omega,A E_{\sf f}\Omega_\omega) \quad \ \ (A \in \mathcal{N}_\omega)\,.
\end{align}
Noting that $E_{\sf f}\Omega_\omega$ is a cyclic and separating unit vector for $\mathcal{N}_\omega$, the previous relations allow  to conclude that
\begin{align}
 S(\omega_{E_{\sf f}\Omega_\omega} \| \omega) & = i \left. \frac{d}{dt} \right|_{t = 0}(E_{\sf f}^*\Omega_\omega,\Delta^{it}E_{\sf f}^*\Omega_\omega) \\
 & = i \left. \frac{d}{dt} \right|_{t = 0}((\cos(1){\bf 1} - i \sin(1) F)\Omega_\omega,\Delta^{it}(\cos(1){\bf 1} - i \sin(1) F)\Omega_\omega) \\
 & = i \left. \frac{d}{dt} \right|_{t = 0} \sin^2(1)(F\Omega_\omega,\Delta^{it}F\Omega_\omega) = \sin^2(1) S(\omega_{\sf f} \| \omega)
\end{align}
where we made use of $\Delta^{it} \Omega_\omega = \Omega_\omega$. 

In comparison, the states $\omega_{E_{\sf f}\Omega_\omega}$ would appear as the counterparts, with respect to a standard state $\omega$ on {\tt CAR}$(\mathcal{K},\Gamma)$,
of the coherent states for the Weyl-algebra of the canonical commutation relations discussed briefly in the Introduction. However, as the considerations just presented
show, their relative entropy $S(\omega_{E_{\sf f}\Omega_\omega} \| \omega)$ with respect 
to a chosen standard state $\omega$ differs from the relative entropy $S(\omega_{\sf f}\|\omega)$ of the single-excitation state $\omega_{\sf f}$ only by the universal numerical 
factor $\sin^2(1)$. In other words, one may equally well regard the single-excitation states $\omega_{\sf f}$ as the counterparts of coherent states on {\tt CAR}$(\mathcal{K},\Gamma)$ in the discussion of relative entropy. 

\section{Relative entropy with respect to quasifree S1PFS}

Any state $\omega$ on {\tt CAR}$(\mathcal{K},\Gamma)$ is completely determined 
by  the collection of its {\it $n$-point functions} as $n$ ranges over $\mathbb{N}$,
\begin{align}
 W_\omega^{(n)}({\sf f}_1,\ldots,{\sf f}_n) = \omega(B({\sf f}_1) \cdots B({\sf f}_n))\,.
\end{align}
Quasifree states on {\tt CAR}$(\mathcal{K},\Gamma)$ are completely determined by their
$2$-point functions, see {\rm \cite{ArakiQuasifreeCAR}}, as follows.
\begin{defn}   
 A state $\omega$ on {\tt CAR}$(\mathcal{K},\Gamma)$ is called \emph{quasifree} if,
 for any $n \in \mathbb{N}$,
\begin{align} \label{eq:np-odd}
 W_\omega^{(n)}({\sf f}_1,\ldots,{\sf f}_{n}) &  = 
                                               0 \quad \ {\rm if}\ n\  {\rm is\ odd}\,,  
                                               \quad {\rm and} \\
                                               \label{eq:np-even}
  W_\omega^{(n)}({\sf f}_1,\ldots,{\sf f}_{n}) &  =                                             
                                              (-1)^{m(m-1)/2} \sum_{p \in \mathcal{P}_{\widehat{2m}}} {\rm sgn}(p)
                                              \prod_{j = 1}^m
                                              W^{(2)}_\omega({\sf f}_{p(j)},{\sf f}_{p(j + m)}) \quad \ {\rm if}\ n =2m\  {\rm is\ even}
\end{align}
where $\mathcal{P}_{\widehat{2m}}$ denotes the set of all permutations 
$p$ of the set $\{1,2,\ldots,2m\}$ which fulfill the conditions $p(1) < \cdots < p(m)$
and $p(j) < p(j + m)$ $(j = 1,\ldots,m)$. 
\end{defn}
Following \cite{Koshida}, it is helpful to note that therefore, 
for a quasifree state $\omega$, the $2m$-point functions are equal to the 
{\it Pfaffian determinant}, or simply {\it Pfaffian} ${\rm Pf}({\bf A})$, of the anti-symmetric 
$2m \times 2m$ matrix ${\bf A} = ( a_{j,k})_{j,k = 1}^{2m}$ whose upper-diagonal
entries are given by 
\begin{align}
 a_{j,k} = W^{(2)}_\omega({\sf f}_j,{\sf f}_k)  \quad \ \ (1 \le j < k \le 2m)\,.
\end{align}
In other words, in the notation of \cite{Vein-Dale},
\begin{align}
W_\omega^{(2m)}({\sf f}_1,\ldots,{\sf f}_{2m}) 
& =  {\rm Pf}({\bf A}) \\[6pt]
& = \quad \left. \begin{array}{cccc}
         \left| \ \, W^{(2)}_\omega({\sf f}_1,{\sf f}_2) \right. & 
         W^{(2)}_\omega({\sf f}_1,{\sf f}_3) &
          \ldots & W^{(2)}_\omega({\sf f}_1,{\sf f}_{2m}) \\
           & W^{(2)}_\omega({\sf f}_2,{\sf f}_3) & \ldots &
           W^{(2)}_\omega({\sf f}_2,{\sf f}_{2m}) \\
           & & & \cdot \\
           & & & \cdot \\
           & & & \cdot \\
           & & & W^{(2)}_\omega({\sf f}_{2m - 1} , {\sf f}_{2m})
        \end{array}
\right| \nonumber
\end{align}
for a quasifree state $\omega$.\footnote{For quasifree states on the {\it non-self-dual} 
CAR algebra, the $n$-point functions for even $n$ can be expressed as a determinant, as e.g.\ in \cite{PowersStoermer}}
\\[6pt]
It is worth mentioning the following property
of the Pfaffian:
\begin{align}
 {\rm Pf}({\bf A})^2 = {\rm det}({\bf A})
\end{align}
for all (complex-valued) anti-symmetric  $2m \times 2m$ matrices ${\bf A}$ \cite{Vein-Dale}. Using this, it is easy to show that,
for any differentiable function $t \mapsto {\bf A}(t)$ $(t \in \mathbb{R})$ valued 
in the anti-symmetric, invertible  $2m \times 2m$ matrices, one obtains
\begin{align}
 \frac{d}{dt} {\rm Pf}({\bf A}(t)) = \frac{1}{2}{\rm Pf}({\bf A}(t))\,{\rm Tr}\left(
 {\bf A}(t)^{-1}\frac{d}{dt}{\bf A}(t)  \right)
\end{align}
where ${\rm Tr}$ denotes the trace. 
\\[6pt]
We can make use of these properties for our next result. We consider a quasifree S1PFS
 $\omega$ on {\tt CAR}$(\mathcal{K},\Gamma)$, where 
the 1-particle unitary group on $\mathcal{K}$ is $\{u_t\}_{t \in \mathbb{R}}$
with $u_t = {\rm e}^{-it\boldsymbol{h}}$.
Employing
again the notation as before, for ${\sf f}_j$ $(j = 1,\ldots,m)$ in $\mathcal{K}$,
with $\Gamma {\sf f}_j = {\sf f}_j$ and $\braket{{\sf f}}{{\sf f}}_{\mathcal{K}} = 2$
we write $F_j = \pi_\omega(B({\sf f}_j))$ in the GNS representation $(\mathcal{H}_\omega,\pi_\omega,\Omega_\omega)$. We will also write ${\sf f}_j(t) = u_t{\sf f}_j$, and 
we note that hence, setting $F_j(t) = \pi_\omega(B({\sf f}_j(t)))$, we have $F_j(t) = \Delta^{it} F_j \Delta^{-it}$. 
\\[6pt]
Our aim is to find a simple expression for 
\begin{align}
 S(\omega_{ {\sf f}_1\cdots {\sf f}_m} \| \omega) &  = i \left. \frac{d}{dt} \right|_{t = 0} (\Omega_\omega,F_1 \cdots F_m \Delta^{it} F_m \cdots F_1 \Omega_\omega) \\
 & = i \left. \frac{d}{dt} \right|_{t = 0} (\Omega_\omega,F_1 \cdots F_m F_m(t)\cdots 
 F_1(t) \Omega_\omega) \,.
\end{align}
Since $\omega$ is quasifree,
we see that 
\begin{align}
& (\Omega_\omega,F_1 \cdots F_m F_m(t)\cdots 
 F_1(t) \Omega_\omega) 
 = W_\omega^{(2m)}({\sf f}_1,\ldots,{\sf f}_m,
 {\sf f}_m(t),\ldots, {\sf f}_1(t)) \\
 & = {\rm Pf}({\bf A}(t)) \label{eq:Bt}
\\[6pt]
& = \ \ \left.
 \begin{array}{crr}
         \left| \ \, W^{(2)}_\omega({\sf f}_1,{\sf f}_2) \right. & 
         W^{(2)}_\omega({\sf f}_1,{\sf f}_3)
       \ \ \ldots \ \
         W^{(2)}_\omega({\sf f}_1,{\sf f}_m) & 
         W^{(2)}_\omega({\sf f}_1,{\sf f}_m(t)) 
      \ \  \ldots \ \
         W^{(2)}_\omega({\sf f}_1,{\sf f}_1(t)) \\
               &   
        W^{(2)}_\omega({\sf f}_2,{\sf f}_3) 
      \  \ \ldots \ \
         W^{(2)}_\omega({\sf f}_2,{\sf f}_m) & 
         W^{(2)}_\omega({\sf f}_2,{\sf f}_m(t)) 
    \  \   \ldots  \ \
         W^{(2)}_\omega({\sf f}_2,{\sf f}_1(t)) \\
          &  \cdot & \\
           &  \cdot & \\
          &  \cdot & \\
          & &  W^{(2)}_\omega({\sf f}_{2}(t) , 
         {\sf f}_1(t))
        \end{array} \right| \nonumber
\end{align}
It should be clear that the entries ${\sf f}_m(t),\ldots,{\sf f}_1(t)$ play the role of entries with lower index labels running from
$m + 1$ to $2m$ in the $2m$-point function, and correspondingly in the Pfaffian. Note that in Equation \eqref{eq:Bt}, the matrix ${\bf A}(t)$
is meant to be the anti-symmetric $2m \times 2m$ matrix whose 
upper diagonal entries are shown in the display array. Since
$W^{(2)}_\omega({\sf f}_j(t),{\sf f}_k(t)) = W^{(2)}_\omega({\sf f}_j,{\sf f}_k) = \braket{{\sf f}_j}{Q_{(1)}{\sf f}_k}_{\mathcal{K}} $, it holds that 
\begin{align}
\frac{1}{i} \left.\frac{d}{dt}\right|_{t = 0} {\bf A}(t) = \left(
 \begin{array}{cc}
  0 & {\bf a} \\
  -{\bf a}^T & 0
 \end{array}
\right)
\end{align}
where the $m \times m$ matrix ${\bf a}$ is given by 
\begin{align} \label{eq:the_b}
 {\bf a} = \left( \begin{array}{ccc}
 \braket{{\sf f}_1}{Q_{(1)}\boldsymbol{h}{\sf f}_m}_{\mathcal{K}} &
 \ldots & \braket{{\sf f}_1}{Q_{(1)}\boldsymbol{h}{\sf f}_1}_{\mathcal{K}} \\
 & \cdot & \\
 & \cdot & \\
 \braket{{\sf f}_m}{Q_{(1)}\boldsymbol{h}{\sf f}_m}_{\mathcal{K}} &
 \ldots & \braket{{\sf f}_m}{Q_{(1)}\boldsymbol{h}{\sf f}_1}_{\mathcal{K}}
 \end{array} \right) 
\end{align}
with the same notation as in \eqref{eq:theQ1}.
On writing ${\bf A} = {\bf A}(t = 0)$, and observing that 
${\rm Pf}({\bf A}) = 1$ since ${\rm Pf}({\bf A}) = (\Omega_\omega,F_1 \cdots F_m F_m \cdots F_1 \Omega_\omega)$ where the $F_j$ are symmetric and unitary, 
we arrive at the following result.
\begin{thm}
 Suppose that $\omega$ is a quasifree {\rm S1PFS} and denote by
 $\boldsymbol{h}$ the self-adjoint generator of the 1-particle flow 
 $\{u_t\}_{t \in \mathbb{R}}$. The
 entropy of the multi-excitation state $\omega_{{\sf f}_1 \cdots {\sf f}_m}$ relative 
 to $\omega$ is given by 
 \begin{align} \label{eq:multient}
 S(\omega_{{\sf f}_1 \cdots {\sf f}_m}\|\omega) = 
 i\left.\frac{d}{dt}\right|_{t = 0} {\rm Pf}({\bf A}(t))
=
 \frac{1}{2}\,{\rm Tr}\left({\bf A}^{-1} \left(\begin{array}{cc}
  0 & -{\bf a} \\
  {\bf a}^T & 0
 \end{array}
\right)
    \right) 
 \end{align}
 where the $m \times m$ matrix ${\bf a}$ is defined in Equation \eqref{eq:the_b}, and ${\bf A}$ is 
 the anti-symmetric $2m \times 2m$ matrix whose upper diagonal part is presented in the 
 scheme \eqref{eq:Bt} for $t = 0$. Moreover, 
 $S(\omega_{{\sf f}_1 \cdots {\sf f}_m}\|\omega)$ is finite if every ${\sf f}_j$ is contained in the domain of $|\boldsymbol{h}|^{1/2}$. 
\end{thm}
Under additional assumptions on the ${\sf f}_1,\ldots,{\sf f}_m$, the 
expression for the relative entropy of a multi-excitation state with respect to 
a quasifree state can still take a much simpler form.
\begin{prop} \label{prop:summation}
 We assume that $\omega$ is a quasifree {\rm S1PFS}, and that 
 we have a multi-excitation state $\omega_{{\sf f}_1 \cdots {\sf f}_m}$ where the 
 ${\sf f}_j$ form an orthonormal family with respect to the 2-point function, i.e.\
\begin{align}
 W^{(2)}_\omega({\sf f}_j,{\sf f}_k) = \delta_{jk}\,, \quad \ \ (j,k = 1,\ldots,m)\,.
\end{align}
Then, 
\begin{align}
 S(\omega_{{\sf f}_1 \cdots {\sf f}_m}\|\omega) =
  \sum_{j = 1}^m S(\omega_{{\sf f}_j}\| \omega) \,.
\end{align}
\end{prop}
{\it Proof}. Under the stated assumptions, we observe that 
\begin{align}
 {\bf A} = \left( \begin{array}{cc}
                   0 & \check{{\bf I}} \\
                   -\check{{\bf I}} & 0
                  \end{array} \right)
\end{align}
where $\check{{\bf I}}$ is the $m \times m$ matrix which has entries 1 on its anti-diagonal, and otherwise zeros:
\begin{align}
\check{{\bf I}} = \left( \begin{array}{ccccc}
                        0 & 0 &\ldots & 0 & 1 \\
                        0 & 0 & \ldots & 1 & 0 \\
                        \cdot & & & & \cdot \\
                         \cdot & & & & \cdot \\
                        1 & 0 & \ldots & 0 & 0
                       \end{array} \right)
\end{align}
Then $\check{{\bf I}}$ coincides with its inverse and we have that 
\begin{align}
 {\bf A}^{-1} = \left( \begin{array}{cc}
                   0 & -\check{{\bf I}} \\
                   \check{{\bf I}} & 0
                  \end{array} \right)  = -{\bf A}\,.
\end{align}
Hence, 
\begin{align} \label{eq:trace-invert}
 S(\omega_{{\sf f}_1 \cdots {\sf f}_m}\|\omega) =
 \frac{1}{2} {\rm Tr}\left( \begin{array}{cc}
        - \check{{\bf I}} {\bf a}^T & 0 \\
         0 & - \check{{\bf I}} {\bf a}
        \end{array} \right) \,.
\end{align}
We see that ${\bf a}$ has the entries $\braket{{\sf f}_1}{Q_{(1)}\boldsymbol{h}{\sf f}_1}_{\mathcal{K}},\ldots,\braket{{\sf f}_m}{Q_{(1)}\boldsymbol{h}{\sf f}_m}_{\mathcal{K}}$ on its anti-diagonal and therefore, 
$\check{{\bf I}} {\bf a}$ has these entries on its diagonal. The same holds for 
$\check{{\bf I}}{\bf a}^T$ in place of $\check{{\bf I}}{\bf a}$. 
We conclude that the expression on the right hand side of \eqref{eq:trace-invert} 
equals $\sum_{j = 1}^m \braket{{\sf f}_j}{Q_{(1)}\boldsymbol{h}{\sf f}_j}_{\mathcal{K}}$. By part (A) of Thm.\ \ref{thm:AandB}, this coincides with
$\sum_{j = 1}^m S(\omega_{{\sf f}_j} \| \omega)$, proving the statement. 
${}$ \hfill $\Box$
\\[6pt]
In view of the fact that the ${\sf f}_j$ $(j = 1,\ldots,m)$ are required to fulfill
also the conditions $\Gamma {\sf f}_j = {\sf f}_j$ and $\braket{{\sf f}_j}{{\sf f}_j}_{\mathcal{K}} = 2$ in order to induce multi-excitation states with respect to $\omega$, the question arises if there actually are collections ${\sf f}_1,\ldots,{\sf f}_m$ with $m >1$ fulfilling the assumptions of the just stated Proposition. 
We   show next that this is indeed the case (barring the trivial
case that $\mathcal{K}$ is 1-dimensional). 
\begin{prop}
 Consider a 1-particle flow
  $\{u_t\}_{t \in \mathbb{R}}$ inducing a 
  1-parameter automorphism group on 
  {\tt CAR}$(\mathcal{K},\Gamma)$, with selfadjoint generator $\boldsymbol{h}$ and 
  associated projector-valued spectral measure $J \mapsto E(J)$.
  Let $\{{\sf e}_j\}_{j \in M}$, with $M \subset \mathbb{N}$, be a finite or
 infinite sequence of unit vectors in $E(J_j)\mathcal{K}$, where the $J_j \subset
 (0,\infty)$ are open intervals which are pairwise disjoint, $J_j \cap J_i = \emptyset$ for $j \ne i$. 
 \\[6pt]
 Then, the collection of ${\sf f}_j = {\sf e}_j + \Gamma {\sf e}_j$ $(j \in M)$ 
 has the following properties:
 \begin{itemize}
  \item[$(1)$] $\Gamma {\sf f}_j = {\sf f}_j$ and $\braket{{\sf f}_j}{{\sf f}_j}_{\mathcal{K}} = 2$ \quad \ \ $(j \in M)$
  \item[$(2)$] $\braket{{\sf f}_j}{Q_{(\beta)}{\sf f}_k}_{\mathcal{K}} = \delta_{jk}$
 \quad \ \  $(j,k \in M)$ \quad \quad
  whenever $\beta > 0$.
 \end{itemize}

\end{prop}
{\it Proof}. The property $\Gamma {\sf f}_j = {\sf f}_j$ is obvious from the 
definition of the ${\sf f}_j$. Since the intervals $J_j$ are pairwise disjoint
and the ${\sf e}_j$ are assumed to be unit vectors in $E(J_j)\mathcal{K}$,
the collection $\{{\sf e}_j\}_{j \in M}$ forms an orthonormal system,
$\braket{{\sf e}_j}{{\sf e}_k}_{\mathcal{K}} = \delta_{jk}$. This follows since
the subspaces $E(J_j)\mathcal{K}$ and $E(J_k)\mathcal{K}$ of $\mathcal{K}$  are orthogonal if $j \ne k$ by the properties of any spectral measure. Clearly, then
also $\{\Gamma{\sf e}_j\}_{j \in M}$ forms an orthonormal system, and since
$\Gamma E(J_j) \Gamma = E(-J_j)$, one concludes that $\braket{\Gamma{\sf e}_j}{{\sf e}_k}_{\mathcal{K}} = 0$ for all $j,k \in M$. Therefore,
\begin{align}
 \braket{{\sf f}_j}{{\sf f}_j}_{\mathcal{K}} 
 & = \braket{{\sf e}_j + \Gamma{\sf e}_j}{{\sf e}_j + \Gamma {\sf e}_j}_{\mathcal{K}} = \braket{{\sf e}_j}{{\sf e}_j}_{\mathcal{K}} + \braket{\Gamma{\sf e}_j}{\Gamma{\sf e}_j}_{\mathcal{K}} = 2
\end{align}
Similarly, using that the spectral projections $E(J_j)$ commute with 
$Q_{(\beta)}$, we find 
\begin{align}
\braket{{\sf f}_j}{Q_{(\beta)}{\sf f}_k}_{\mathcal{K}} 
 & = \braket{{\sf e}_j + \Gamma{\sf e}_j}{Q_{(\beta)}({\sf e}_k + \Gamma {\sf e}_k)}_{\mathcal{K}} = \braket{{\sf e}_j}{Q_{(\beta)}{\sf e}_k}_{\mathcal{K}} + \braket{\Gamma{\sf e}_j}{Q_{(\beta)}\Gamma{\sf e}_k}_{\mathcal{K}} = \delta_{jk}
\end{align}
on recalling that $Q_{(\beta)} + \Gamma Q_{(\beta)} \Gamma = {\bf 1}$.
This completes the proof. \hfill $\Box$

\section{Comparison with relative entropy defined using von Neumann entropy}

In this section we will demonstrate explicitly that the 
relative entropy of multi-excitation states $\tilde{\omega} = \omega_{{\sf f}_1 \cdots {\sf f}_n}$ with respect to a quasifree S1PFS $\omega$ on {\tt CAR}$(\mathcal{K},\Gamma)$ coincides with the 
 relative entropy obtained if the states are represented as 
density matrices $\varrho_{\tilde{\omega}}$ and $\varrho_\omega$ with respect to a given, irreducible Hilbert space representation, as given by the familiar formula,
\begin{align}
 S(\tilde{\omega} \|\omega) = {\rm Tr}(\varrho_\omega(\log(\varrho_\omega) - \log(\varrho_{\tilde{\omega}}))) \,.
\end{align}
In the present case, the given Hilbert space representation is a 
quasifree ground state. While the result is quite well-known in the case where $\mathcal{K}$ is finite-dimensional (see, e.g., \cite{OhyaPetz-book}), we think that it will be instructive to present it also in order
to clarify the connection of the abstractly described quasifree S1PFS with their more familiar description as density matrix states in a ground state Hilbert space representation -- provided that the quasifree standard states arise in that way.
\\[6pt]
We   begin the discussion by   imposing the condition of ``absence of zero modes for 
$\boldsymbol{h}$'', where $\boldsymbol{h}$ is the selfadjoint generator of a one-particle flow on
$(\mathcal{K},\Gamma)$. 
Thus,  the
spectral projector of the positive real axis,  $P = E((0,\infty))$, is a base polarizator, and it defines a quasifree ground state for 
the automorphism group $\{ {\tt T}_t\}_{t \in \mathbb{R}}$ induced by the 1-particle flow \cite{ArakiQuasifreeCAR}. (We will summarize the definition of ground state in the Appendix.)

The  quasifree ground state $\omega^{P}$ is 
defined by the 2-point function 
\begin{align}
W^{(2)}_{P}({\sf f},{\sf g}) = \braket{\Gamma {\sf f}}{P{\sf g}}_{\mathcal{K}}
\end{align}
The 1-particle flow $\{u_t\}_{t \in \mathbb{R}}$ commutes with $\Gamma$ and with $P$, implying $W^{(2)}_{P}(u_t{\sf f},u_t{\sf g}) = W^{(2)}_{P}({\sf f},{\sf g})$ for all real $t$ and ${\sf f},{\sf g} \in \mathcal{K}$, implying in turn $\omega^{P}({\tt T}_t(A)) = \omega^{P}(A)$ 
$(A \in {\tt CAR}(\mathcal{K},\Gamma))$, i.e.\ the invariance of $\omega^{P}$ under the automorphisms induced by the 1-particle flow. 
\\[6pt]
We denote the GNS representation of $\omega^{P}$ by $(\mathcal{H}^{P},\pi^{P},\Omega^{P})$. This GNS representation is irreducible, i.e.\
the induced von Neumann algebra $\mathcal{N}_{\omega^{P}}$ coincides
with $\mathcal{B}(\mathcal{H}^{P})$ \cite{ArakiQuasifreeCAR}. 
By general arguments, as $\omega^{P}$ is invariant under 
$\{ {\tt T}_t \}_{t \in \mathbb{R}}$,
there is a continuous unitary group $\{U_t\}_{t \in \mathbb{R}}$ on 
$\mathcal{H}^{P}$ so that 
\begin{align}
 U_t \pi^{P}(B({\sf f}))U_t^* = \pi^{P}(B(u_t{\sf f})) \quad \text{and} \quad 
 U_t \Omega^{P} = \Omega^{P} \quad \ \ ({\sf f} \in \mathcal{K}\,,\ t \in \mathbb{R})
\end{align}
We denote the selfadjoint generator of that unitary group by $\boldsymbol{H}$,
$U_t = {\rm e}^{-it\boldsymbol{H}}$. 
\begin{prop}
 Assume that for any $\beta > 0$, ${\rm e}^{-\beta \boldsymbol{H}}$ is a 
 trace-class operator on $\mathcal{H}^{P}$, and define the density
 matrix 
 \begin{align}
  \varrho_\beta = \frac{1}{{\rm Tr}({\rm e}^{-\beta \boldsymbol{H}})} {\rm e}^{-\beta \boldsymbol{H}}
 \end{align}
{\rm (A)} \ \ The state on {\tt CAR}$(\mathcal{K},\Gamma)$ given by 
\begin{align}
 \omega^{(\beta)}(A) = {\rm Tr}(\varrho_\beta \pi^{P}(A)) \quad \ \ 
 (A \in {\tt CAR}(\mathcal{K},\Gamma))
\end{align}
is a KMS state at inverse temperature $\beta$ for $\{ {\tt T}_t \}_{t \in \mathbb{R}}$. 
As the condition of ``absence of zero modes for $\boldsymbol{h}$'' has been imposed, 
$\omega^{(\beta)}$ coincides with the quasifree KMS state defined by the polarizator 
$Q_{(\beta)}$ in Equation \eqref{eq:Qbeta}.
\\[6pt]
{\rm (B)} \ \ Let ${\sf f}_j$, $j = 1,\ldots,n$ be in $\mathcal{K}$, with 
$\Gamma {\sf f}_j = {\sf f}_j$ and $\braket{{\sf f}_j}{{\sf f}_j}_{\mathcal{K}} = 2$.
Then, the operator $F = \pi^{P}(B({\sf f}_1)\cdots B({\sf f}_n))$ on $\mathcal{H}^P$ is unitary.
Defining the density matrices $\varrho = \varrho_{(1)}$ and $\tilde{\varrho} = F \varrho_{(1)} F^*$,
and the corresponding states 
\begin{align}
 \omega(A) = {\rm Tr}(\varrho\pi^P(A)) \quad \text{and} \quad \tilde{\omega}(A)
 ={\rm Tr}(\tilde{\varrho} \pi^P(A)) \quad \ \ (A \in {\tt CAR}(\mathcal{K},\Gamma))\,,
\end{align}
it holds that 
\begin{align} \label{eq:3forms}
 S(\omega_{{\sf f}_1\cdots {\sf f}_n}\|\omega) = S(\tilde{\omega}\|\omega) = 
 {\rm Tr}(\varrho(\log(\varrho) - \log(\tilde{\varrho})))
\end{align}
\end{prop}
{\it Proof}. 
(A) \ \ This is a well-known fact, see e.g. \cite{ArakiQuasifreeCAR,BraRob2}.
\\[6pt]
(B) \ \
Set $B = B({\sf f}_1)\cdots B({\sf f}_n)$. Then 
\begin{align}
\tilde{\omega}(A) & = {\rm Tr}(F \varrho F^*\pi^P(A)) = {\rm Tr}(\varrho F^* \pi^P(A)F) \\
& = 
{\rm Tr}(\varrho \pi^P(B^* A B))
= \omega(B^* A B)
\quad \ \ 
(A \in {\tt CAR}(\mathcal{K},\Gamma))
\end{align}
Therefore, expressed in the GNS representation $(\mathcal{H}_\omega,\pi_\omega,\Omega_\omega)$,
\begin{align}
 \tilde{\omega}(A) = (F_\omega\Omega, \pi_\omega(A) F_\omega\Omega_\omega)
 \quad \ \ 
(A \in {\tt CAR}(\mathcal{K},\Gamma))
\end{align}
where $F_\omega = \pi_\omega(B)$ and consequently,
\begin{align}
 S(\omega_{{\sf f_1} \cdots {\sf f}_n}\|\omega) & = S(\tilde{\omega}\|\omega)  = i\left.\frac{d}{dt}\right|_{t = 0} (\Omega_\omega,F_\omega \Delta^{it}F_\omega^*\Omega_\omega) \\
 & = i\left.\frac{d}{dt}\right|_{t = 0}\omega(B{\tt T}_t(B^*)) 
  = i\left.\frac{d}{dt}\right|_{t = 0}{\rm Tr}(\varrho F U_tF^*U_t^*) \\
 & = {\rm Tr}(\varrho F[\boldsymbol{H},F^*])
\end{align}
On the other hand, one obtains
\begin{align}
 {\rm Tr}(\varrho(\log(\varrho) - \log(\tilde{\varrho}))) = {\rm Tr}(\varrho((-\boldsymbol{H}) -
 F (- \boldsymbol{H})F^*) = {\rm Tr}(\varrho F[\boldsymbol{H},F^*])
\end{align}
proving that all the terms in relation \eqref{eq:3forms} agree. \hfill $\Box$

\section{Conclusion}
We have studied the relative entropy for multi-excitation states with respect to 
S1PFS on the self-dual CAR algebra. The S1PFS $\omega$, which equals a KMS-state at inverse temperature
$\beta = 1$, for a unitary dynamics on the CAR one-particle space, is unique under very 
general circumstances, and quasifree, with 2-point function given by \eqref{eq:unique}. This 2-point function
is entirely determined by the one-particle Hilbert space and the self-adjoint generator 
$\boldsymbol{h}$ of the unitary dynamics on the one-particle space, and likewise is the 
expression for the relative entropy of any single- or multi-excitation state, as stated in
\eqref{eq:1exent} and \eqref{eq:multient}, respectively. This is very similar to the 
expression for the relative entropy of coherent states on the CCR algebra \eqref{eq:coherentent}.
In both cases, there is a preferred ``reference'' state $\omega$ -- the quasifree S1PFS in for the CAR 
algebra -- so that there is, from this perspective, no dependence of the relative entropy on any potential choice for 
$\omega$. For a quasifree S1PFS $\omega$, the fact that the $2m$-point functions are 
given by a Pfaffian determinant whose entries are formed by 2-point functions, allows it 
to derive the compact formula \eqref{eq:multient} for the relative entropy of multi-excitation states. 
\par
One can therefore conclude that the multi-excitation states with respect to S1PFS on the self-dual CAR algebra play a very similar role as coherent states on the CCR-Weyl algebra when discussing relative entropy: Their relative entropies can be explicitly calculated and can be expressed fully by the one-particle Hilbert space and the self-adjoint generator $\boldsymbol{h}$ of the one-particle dynamics. 
\par
Some applications appear in \cite{Galanda}. We mention that one can think of other definitions
of relative entropy for states on the self-dual CAR algebra, e.g.\ by associating them with
coherent states in a ``bosonized'' framework \cite{Combescure-Robert}. It would be interesting to see if 
matching results for the relative entropy ensue. 
${}$ \\ \\ ${}$ \\
{\bf  Acknowledgments}
S.G. would like to thank the ITP in Leipzig for the hospitality in February, the National Group of Mathematical Physics (GNFM-INdAM) for the support, Simone Murro and Nicola Pinamonti for discussions. A.M. would like to thank Markus Fröb and Felix Finster for several discussions. 
\\\\\\\bigskip
{\bf \Large Appendix}
\\ \\
We summarize the definitions of ground state and KMS state with 
respect to an automorphism group $\{{\tt T}_t\}_{t \in \mathbb{R}}$ on
{\tt CAR}$(\mathcal{K},\Gamma)$ from \cite{ArakiQuasifreeCAR}; see also 
\cite{BraRob2} for further discussion.  The automorphism group need not be 
induced by a 1-particle flow, and in fact, the definitions apply, more generally,
for continuous 1-parametric automorphism groups defined on any $C^*$-algebra.
\\[6pt]
A state $\omega$ on {\tt CAR}$(\mathcal{K},\Gamma)$ is called a {\it ground state}
for $\{{\tt T}_t\}_{t \in \mathbb{R}}$ 
if 
\begin{align}
 \int_\mathbb{R}  \chi(t) \omega(A_1 {\tt T}_t(A_2))\,dt = 0 \quad \ \ (A_1,A_2 \in
 {\tt CAR}(\mathcal{K},\Gamma))
\end{align}
for all Schwartz type test functions $\chi$ with ${\rm supp}(\hat{\chi}) \subset 
(-\infty, 0)$, where $\hat{\chi}$ denotes the Fourier transform,
\begin{align}
 \hat{\chi}(s) = \frac{1}{\sqrt{2\pi}} \int_\mathbb{R} \chi(t){\rm e}^{-ist}\,dt \quad \ \ (s \in\mathbb{R}) \,.
\end{align}
A state $\omega$ on {\tt CAR}$(\mathcal{K},\Gamma)$ is called a {\it KMS state at inverse temperature $\beta > 0$}
for $\{{\tt T}_t\}_{t \in \mathbb{R}}$  
if 
\begin{align}
 \int_\mathbb{R} \chi(t) \omega(A_1 {\tt T}_t(A_2)) \,dt = 
  \int_{\mathbb{R}} \chi(t +i\beta) \omega({\tt T}_t(A_2)A_1)\,dt \quad \ \
  (A_1,A_2 \in
 {\tt CAR}(\mathcal{K},\Gamma))
\end{align}
for all test functions $\chi$ with $\hat{\chi} \in C_0^\infty(\mathbb{R})$.
Recall that if $\hat{\chi} \in C_0^\infty(\mathbb{R})$, then $\chi$ is a Schwartz type test function which possesses an analytic extension to the complex plane, by the Paley-Wiener theorem.


\medskip

\printbibliography

\end{document}